\documentclass[article,twocolumn,prb,showpacs,superscriptaddress]{revtex4-2}

\usepackage{graphicx}
\usepackage{amssymb}
\usepackage{bm}
\usepackage{amsmath}
\usepackage{xfrac}
\usepackage{color}
\usepackage{titlesec}
\renewcommand{\figurename}{Figure} 
\makeatletter	        
\def\fnum@figure{\textbf{\figurename~\thefigure}}
\makeatother
\usepackage[normalem]{ulem}
\usepackage[utf8]{inputenc}
\usepackage[T1]{fontenc}
\usepackage[marginal]{footmisc}

\usepackage{latexsym}

\usepackage[colorlinks=True,
                 linkcolor=black,
	        citecolor=blue,
	        urlcolor=blue]{hyperref}

\makeatletter	        
 \def\section{%
  \@startsection{section}{1}{\z@}{0.8cm plus1ex minus.2ex}{0.2cm}%
  {%
   \small\sffamily\bfseries\selectfont
   \raggedright
   \parindent\z@
  }%
 }%
  \def\subsection{%
  \@startsection{subsection}{2}{\z@}{0.8cm plus1ex minus.2ex}{0.2cm}%
  {%
   \small\sffamily\bfseries
   \raggedright
   \parindent\z@
  }%
 }%
\makeatother

\newcommand{\comment}{\textcolor{black}}

\usepackage{color}

\newcommand{\MOE}{MOE Key Laboratory for Nonequilibrium Synthesis and Modulation of Condensed Matter, Shaanxi Province Key Laboratory of Advanced Materials and Mesoscopic Physics, School of Physics, Xi’an Jiaotong University, Xi’an,710049, China}

\newcommand{\QD}{State Key Lab of Metastable Materials Science and Technology, Yanshan University, Qinhuangdao 066004, China}
\newcommand{\HF}{Anhui Key Laboratory of Low-Energy Quantum Materials and Devices, High Magnetic Field Laboratory, Hefei Institutes of Physical Science,Chinese Academy of Sciences, Hefei 230031, China}
\newcommand{\HFF} {Science Island Branch of Graduate School, University of Science and Technology of China, Hefei 230026, China}

\newcommand{\Yanshan}{Key Laboratory for Microstructure Material Physics of Hebei Province, Yanshan University, Qinhuangdao 066004, China}

\makeatletter
\g@addto@macro\bfseries{\boldmath}
\makeatother

\begin{document}
\title{Bias Voltage Controlled Inversions of Tunneling Magnetoresistance in van der Waals Heterostructures 
Fe$_3$GaTe$_2$/hBN/Fe$_3$GaTe$_2$}

\author{Lihao Zhang}
\affiliation{\MOE}
\author{Miao He}
\affiliation{\HF}
\affiliation{\HFF}
\author{Xiaoyu Wang}
\author{Haodong Zhang}
\affiliation{\MOE}
\author{Keying Han}
\affiliation{\QD}
\affiliation{\Yanshan}
\author{Yonglai Liu}
\affiliation{\HF}
\affiliation{\HFF}
\author{Lei Zhang}
\affiliation{\MOE}
\author{Yingchun Cheng}
\affiliation{\QD}
\affiliation{\Yanshan}
\author{Jie Pan}
\email{Corresponding author: Jie Pan, jiepan@xjtu.edu.cn}
\affiliation{\MOE}
\author{Zhe Qu}
\email{Corresponding author: Zhe Qu, zhequ@hmfl.ac.cn}
\affiliation{\HF}
\affiliation{\HFF}
\author{Zhe Wang}
\email{Corresponding author: Zhe Wang, zhe.wang@xjtu.edu.cn}
\affiliation{\MOE}



\begin{abstract}

 We report the bias voltage-controlled inversions of tunneling magnetoresistance (TMR) in magnetic \comment{tunnel} junctions composed of Fe$_3$GaTe$_2$ electrodes and hBN tunneling barrier, observed at room temperature. The polarity reversal of TMR occurs consistently at \comment{around 0.625 V} across multiple devices and temperatures, highlighting the robustness of the effect. To understand this behavior, we developed a theoretical model incorporating spin-resolved density of states (DOS) at high energy levels. By adjusting the DOS weighting at different k-points to account for misalignment between the crystal structure of electrodes in experimental devices, we improved agreement between experimental and theoretical inversion voltages. Our results provide valuable insight into the voltage-controlled spin injection and detection in two-dimensional magnetic tunnel junctions, with implications for the development of energy-efficient spintronic devices.   
 
\end{abstract}

\maketitle

\section*{Introduction}

Magnetic \comment{tunnel} junctions (MTJs) are critical components in spintronic devices \cite{RMP2004,tsymbal2003spin,RMP2020,Review2020}, such as hard disk drive read heads and magnetoresistive random-access memory. An MTJ consists of a tunneling barrier sandwiched between two ferromagnetic electrodes, each with spin polarized density of states (DOS) at the Fermi level. This structure enables substantial tunneling magnetoresistance (TMR) when the relative magnetization of the two ferromagnets switches between parallel and antiparallel configurations. A key advantage of MTJs is their non-volatile nature—the resistance states remain stable without continuous power, making MTJs attractive for emerging memory technologies, including in-memory computing for Artificial Intelligence and the Internet of Things. However, MTJs with conventional thin films will face challenges as devices scale down to meet the requirements of advanced technological nodes\cite{Nature2022Review}, due to issues such as atomic intermixing at interfaces. 

MTJ composed with purely van der Waals (vdW) 2D materials offer a promising alternative. These materials can retain high quality even at atomic thickness and form atomically flat, defect-free interfaces without atomic intermixing\cite{geim2013van}. Since the discovery of 2D magnets\cite{huang2017layer,gong2017discovery,
deng2018gate,fei2018two}, all-vdW MTJs—such as those based on Fe$_3$GeTe$_2$ and hBN—have demonstrated good performance \cite{NL2018}. Various combinations of tunneling barriers \cite{FGT2020MoS2,FGTAM2021,FGTNM2022,FGTNC2023,FGT2024MoSe} and ferromagnetic electrodes \cite{FGaT2022CPL,FGaT2023NanoS,FGaT2023AMI,FGaT2023SB,FGaT2024Info} based on purely 2D materials have been explored to enhance device functionality. 

Among the intriguing phenomena observed in vdW MTJs is the inversion of TMR polarity under controlled bias voltage \cite{FGTNM2022, FGTNC2023, FGaT2023AMI, FGaT2023SB,FGT2024MoSe,FGaT2024Info}, TMR being positive under certain bias voltage and changing to negative under other bias voltages. This effect suggests that bias voltage can regulate spin injection and detection, and holds great potential for energy-efficient spintronic devices. The voltage-controlled TMR inversion in vdW MTJs is generally attributed to high-energy localized spin states in the ferromagnetic electrodes. Previous studies on Fe$_3$GeTe$_2$-based MTJs with hBN, WSe$_{2}$, and GaSe tunneling barriers have quantitively discussed the voltage thresholds for TMR polarity reversal, often finding that experimental results exceed theoretical predictions\cite{FGTNM2022, FGTNC2023}. For MTJs using ferromagnet Fe$_3$GaTe$_2$ (FGT) with Curie temperature above room temperature which has great potential for future applications, TMR polarity reversal were reported with tunneling barrier of semiconducting WS$_{2}$ and WSe$_{2}$, but these studies did not quantitatively address the underlying mechanisms \cite{FGaT2023AMI, FGaT2024Info}. 

\begin{figure*}
\centering
\includegraphics[width =1\linewidth]{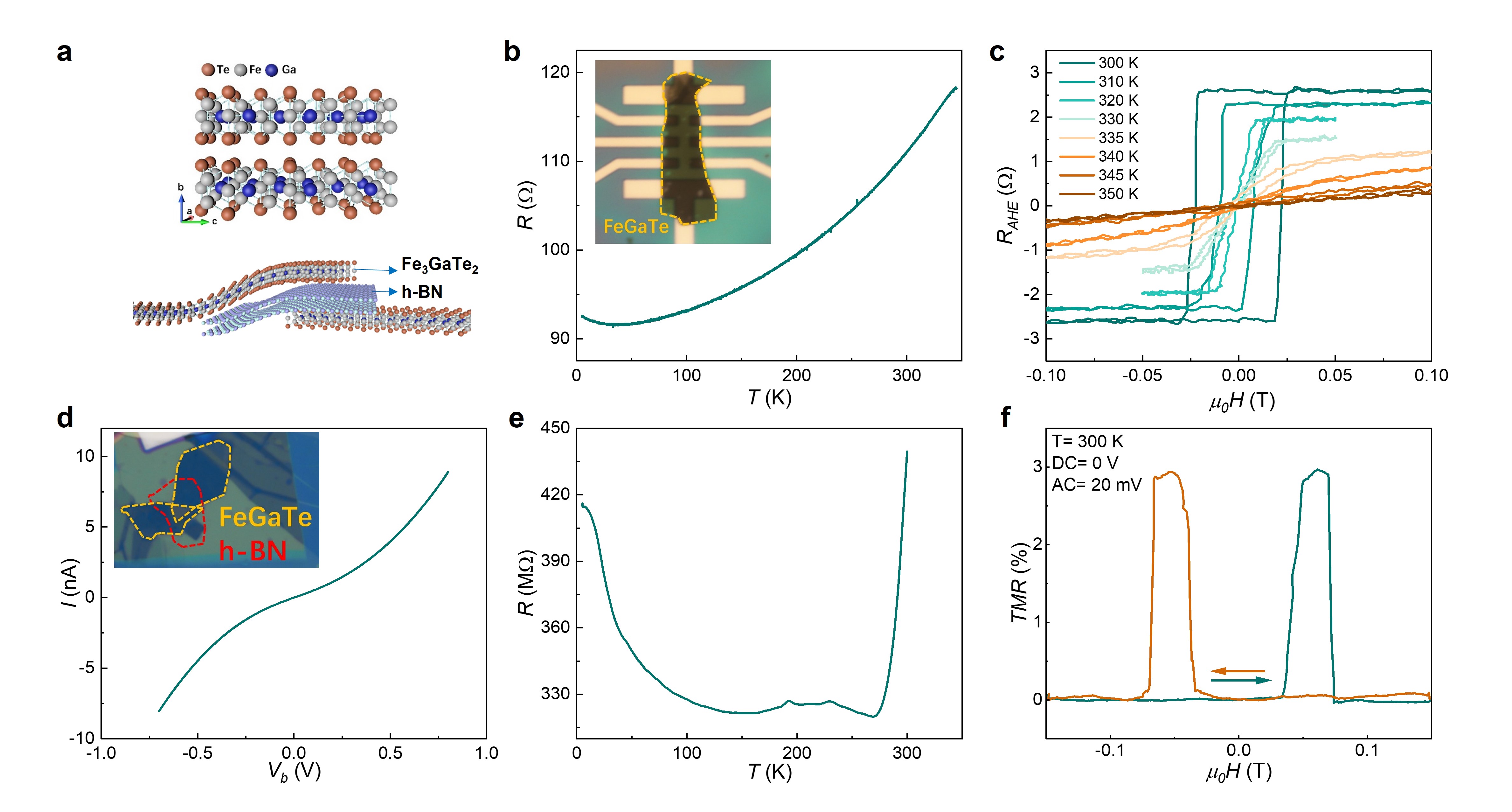}
\caption{\textbf{Basic characteristics of Fe$_3$GaTe$_2$ flakes and magnetic \comment{tunnel} junction (MTJ) of Fe$_3$GaTe$_2$/hBN/Fe$_3$GaTe$_2$.} \textbf{a} Crystal structure of Fe$_3$GaTe$_2$ (up) and schematic of MTJ (down). \textbf{b} Temperature dependence of longitudinal resistance of Fe$_3$GaTe$_2$. The inset show the optical image of the device. \textbf{c} Anomalous Hall effect of Fe$_3$GaTe$_2$ above room temperature. \textbf{d} Voltage dependence of current for \comment{MTJ} at 300 K.  \textbf{e} Temperature dependence of tunneling resistance. \comment{The measurement is done with constant bias voltage of 20 mV and no magnetic field is applied.} \textbf{f} Tunneling magnetoresistance (TMR) of MTJ at 300 K, the applied AC bias is 20 mV and DC voltage is 0 V.} 
\end{figure*}

In this work, we investigate the behavior of MTJs with FGT electrodes and an insulating hBN barrier. \comment{The very large band gap of high quality hBN ensure the electronic transport would be tunneling for the whole investigated temperature range. It also results in large tunneling barrier height so its momentum dependence can be ignored, thus we can focus on the crystal structure alignment of ferromagnetic electrodes.} \comment{On the experimental side,} we demonstrate bias voltage-controlled TMR inversion at room temperature, and show that the voltage threshold for TMR polarity reversal remains consistent across different temperatures and devices. Furthermore, we present theoretical insights that more accurately match the experimental data by applying reduced weighting to the DOS away from the $\Gamma$  point, highlighting the importance of considering in-plane momentum conservation during tunneling and the reality that the crystal lattices of the ferromagnetic electrodes are not aligned.

\section*{Experiments and Results}

We began our experiments by characterizing the thin flakes of FGT, whose crystal structure is shown in Fig. 1a. The single crystals were grown using the chemical vapor transport method \cite{FGaT}, and mechanically exfoliated in the glove box filled with Nitrogen. An optical image of an exfoliated FGT flake on pre-deposited Cr/Au electrodes is shown in the inset of Fig. 1b. We measured both the longitudinal resistance and Hall resistance of the flake with the cryostat from Cryogenic Company, as shown in Figs. 1b and 1c, respectively. The four-probe resistance measurements exhibit metallic behavior, as expected for FGT. The anomalous Hall effect (AHE) shows a characteristic square shape at room temperature, which changes above 320 K due to the formation of magnetic domains \cite{FGTDomain}. The AHE disappears around 340 K, confirming the high quality of our FGT crystals and their Curie temperature.

We assembled a van der Waals heterostructure FGT/hBN/FGT using a standard pick-up technique with PDMS/PC stamps inside a glove box. The inset of Fig. 1d displays an optical image of the fabricated device, in which approximately 4 layers of hBN were used as the tunneling barrier. Fig. 1d shows the voltage dependence of the current measured at 300 K. The I-V curve is slightly non-linear at high voltages, \comment{being} consistent with the expected behavior of \comment{tunnel} junctions. The resistance only change slightly \comment{(around 20\%) during the cool down process}, further confirming the tunneling nature of the electronic transport in the device. Fig. 1f presents the TMR when a magnetic field is applied along the c-axis of the FGT. Due to differences in the thickness and shape of the two FGT electrodes, the coercive fields of the two ferromagnets differ. This results in an antiparallel magnetization configuration over a certain magnetic field range, while a parallel configuration is observed in the remaining field range. Typical spin-valve TMR behavior is thus clearly evident at 300 K.

\begin{figure}
\includegraphics[width =1\linewidth]{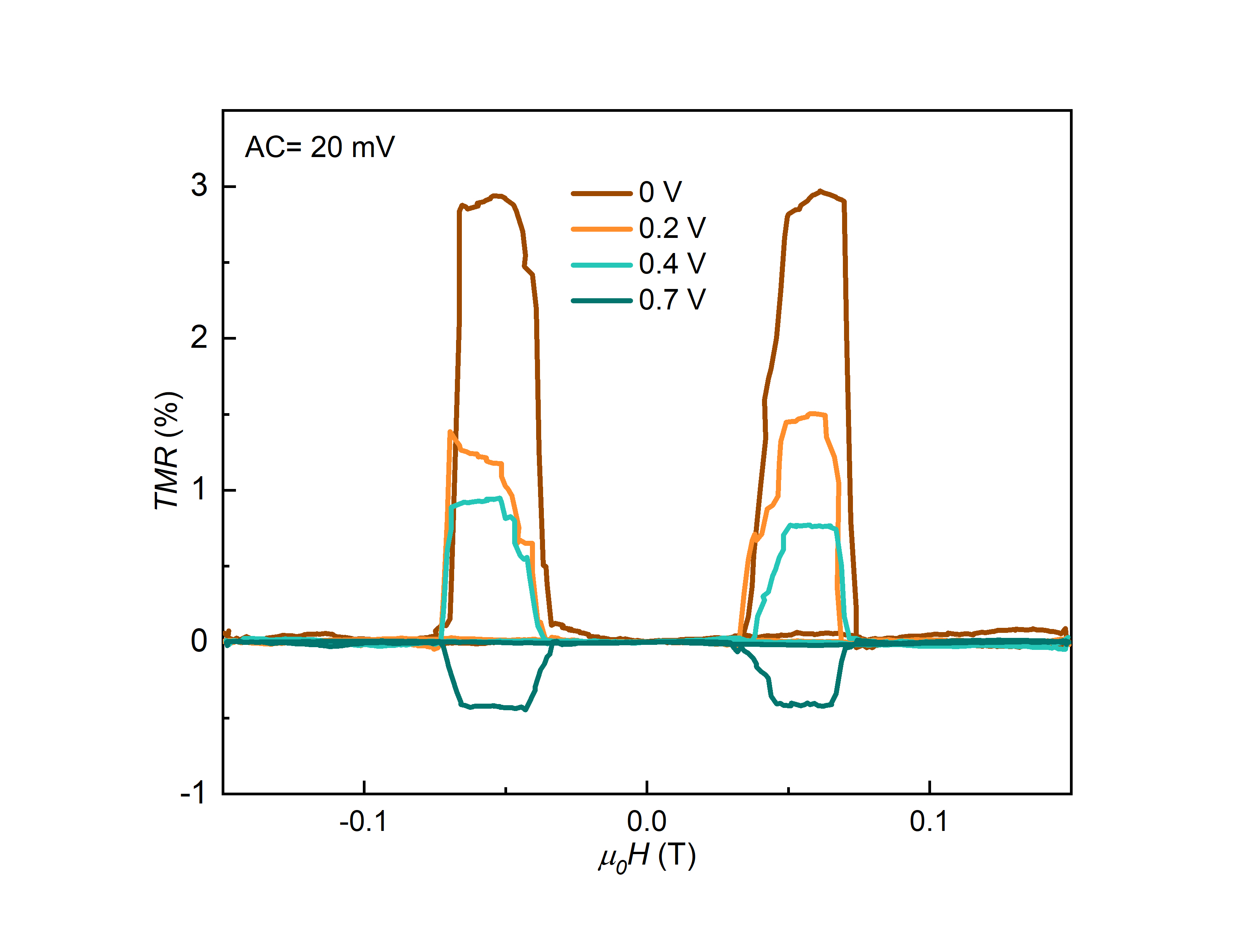}
\caption{\textbf{Bias voltage controlled inversion of TMR at 300 K}. TMR at 300 K with different DC bias voltages.} 
\end{figure}

Fig. 2 shows the TMR measured at 300 K under different bias voltages. While the magnetic field corresponding to the tunneling resistance jumps remains constant, the amplitude of the TMR decreases with increasing bias voltage. Notably, the TMR value eventually becomes negative, meaning the tunneling resistance in the antiparallel configuration becomes smaller than that in the parallel configuration. This observation goes beyond the predictions of the simple Julliere model \cite{julliere} \comment{TMR=2P1$\times$P2/(1-P1$\times$P2) with P1 (P2) being the spin polarization of top (bottom) ferromagnet,} which would typically predict a positive TMR when both electrodes are made of the same material with the same spin polarization. \comment{We discuss the origin of this phenomenon later.}



To further understand the device behavior, we measured the TMR at different temperatures, with the zero-bias data presented in Fig. 3a. The spin-valve effect becomes almost indiscernible at 310 K, which could be due to two factors: either the FGT flakes in this device have a relatively lower Curie temperature, or the formation of complicated magnetic domains above 310 K in the FGT electrodes is suppressing the spin-valve behavior. As the temperature decreases, the tunneling resistance jumps shift to higher magnetic fields, \comment{being} consistent with the expected increase in coercive field at lower temperatures for ferromagnetic metals. Simultaneously, the amplitude of the TMR increases as the temperature drops.

We extracted the spin polarization of the FGT electrodes at zero-bias voltage, \comment{which is the same for top and bottom electrode and result in} TMR=$2P^2/(1-P^2)$. The temperature dependence of the spin polarization is plotted in Fig. 3b, assuming a Curie temperature (T$_c$) of 310 K for the FGT in the MTJ device. The data reveals that the spin polarization increases as the temperature decreases. The temperature dependence of spin polarization, as detected through TMR, is proportional to the magnetization at the surface of the material. In 2D materials, this relationship has been shown to correlate well with bulk magnetization measurements obtained through the AHE \cite{NL2018}. The anomalous Hall resistance can be described by the empirical formula R$_{xy}$=R$_s$M, where M is the material's magnetization and R$_s$ is the anomalous Hall coefficient, which is linked to the longitudinal resistivity and varies with the mechanism of the AHE in the system \cite{AHERMP}. Recent studies have shown that the AHE in FGT arises from an intrinsic mechanism \cite{AHE2024CHX, AHE2024}, where R$_s$ is proportional to R$_{xx}^2$. Consequently, the anomalous Hall conductance, G$_{xy}^A$=R$_{xy}$/(R$_{xx}^2$+R$_{xy}^2$ )$\approx$ R$_{xy}$/R$_{xx}^2$, is proportional to the spontaneous magnetization. We calculated G$_{xy}^A$ from the longitudinal resistance and AHE measurements, and its temperature dependence is displayed in Fig. 3b. An excellent agreement between the spin polarization and the measured anomalous Hall conductance is observed in our MTJ. 

\begin{figure*}
\centering
\includegraphics[width =1\linewidth]{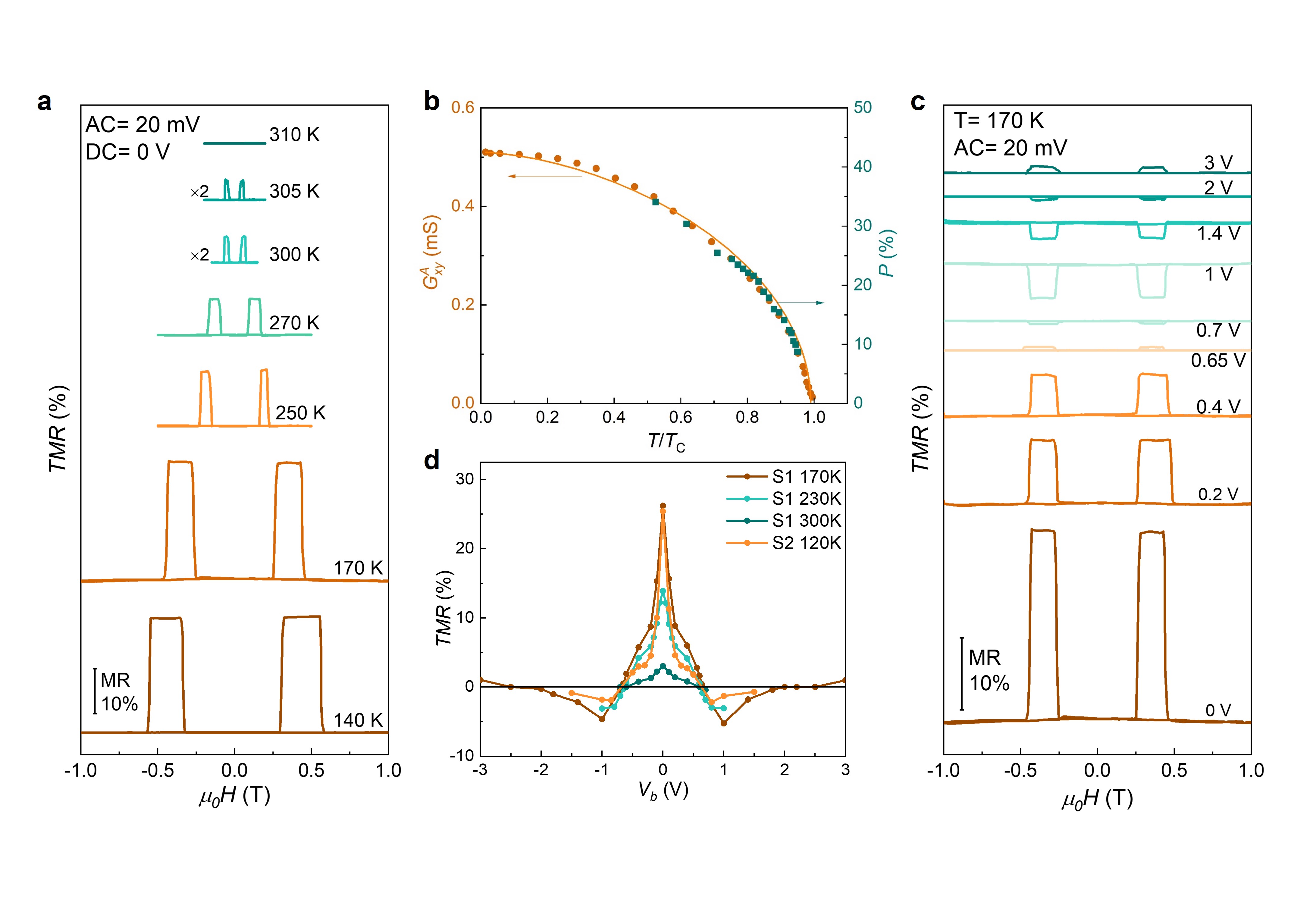}
\caption{\textbf{Bias voltage controlled inversion of TMR at different temperatures and different devices.} \textbf{a} TMR at different temperature with zero voltage bias. \textbf{b} Temperature dependence of anomalous Hall conductance (left axis) and spin polarization (right axis).\comment{The solid orange line is a fit of G$_{xy}^A$ with function commonly used for temperature dependence of magnetization M(T)=M(0)(1-(T⁄T$_c$ )$^\alpha$)$^\beta$, with $\alpha$=1.65 and $\beta$=0.5}. \textbf{c} TMR at 170 K with different bias voltages. \textbf{d} Bias voltage dependence of TMR for \comment{different} temperatures and another device S2. \comment{TMR sign changes at 0.625±0.025 V and 2.5±0.5 V.}} 
\end{figure*}

To model the temperature dependence of the spin polarization and anomalous Hall conductance, which is proportional to the spontaneous magnetization, we applied a commonly used empirical function for magnetization \cite{MTfit}: M(T)=M(0)(1-(T⁄T$_c$ )$^\alpha$)$^\beta$. This interpolation between the Bloch T$^{3/2}$ law at low temperatures and the critical behavior near the Curie temperature includes the critical exponent $\beta$ and $\alpha$ as a fitting parameter. As shown in Fig. 3b, the experimental data fits well when $\beta$=0.5, which corresponds to the mean field theory. We notice that this temperature dependence of anomalous Hall conductance also matches the Brillouin function, \comment{being} consistent with critical exponent obtained above. However, this is different from the critical exponent value of 0.35 \cite{MTZXX} or 0.4 \cite{MTACS}, obtained from bulk magnetization analysis using the Kouvel-Fisher method and modified Arrott plot . While the origin of this discrepancy requires further investigation, it is noteworthy that our anomalous Hall data aligns with previously reported behavior of FGT \cite{AHE2024CHX,AHE2024}. The anomalous Hall conductance in FGT remains almost constant at low temperatures until about 0.4T$_c$, contrasting with Fe$_3$GeTe$_2$, where it decreases nearly linearly with temperature up to 0.8T$_c$ \cite{NL2018}. 

A larger spin polarization results in a higher TMR at lower temperatures, allowing us to study the bias voltage dependence of TMR over a wider range. In Fig. 3c, we show the TMR measured at 170 K. The TMR is positive at zero bias and switches to negative \comment{at 0.625±0.025} V, being consistent with the behavior observed at 300 K. As the bias voltage increases further, the TMR amplitude grows and reaches a maximum at approximately 1 V before decreasing again. Interestingly, the TMR reverts back to a positive value at a bias voltage of around 3 V. 

The overall bias voltage dependence of TMR is summarized in Fig. 3d, which includes data from the primary device in the main text and another device shown in the supplementary materials. The TMR decreases as the bias increases from 0 V, transitioning from positive to negative at \comment{0.625±0.025} V, regardless of temperature or device. While the anomaly at zero bias could be explained by the excitation of magnons due to finite voltage energy \cite{TMRMagnon}, such an effect would only reduce spin polarization without reversing the TMR polarity. Therefore, this explanation cannot account for the TMR reversal observed in our measurements. Interfacial states between the ferromagnetic electrodes and the tunneling barrier have been proposed to explain similar TMR reversals in conventional thin-film MTJs \cite{TMRinterface,TMRinterPRB,TMRinter2013,TMRinter2014,TMRinter2014APL}. However, this mechanism does not apply to our van der Waals heterostructures, as no chemical bonds exist at the surfaces of FGT and hBN. Another proposed mechanism involves the interference of wavefunctions within the tunneling barrier \cite{TMROscillation1,TMROscillation}, which could lead to TMR inversions at high bias voltages. However, this explanation requires transport in the Fowler-Nordheim (FN) tunneling regime. We verified the I-V characteristics of our device by plotting ln(I/V$^2$) against 1/V, confirming that our device does not enter the FN regime until 3 V, ruling out this mechanism in our device for lower voltages. 

\begin{figure*}
\includegraphics[width =0.9\linewidth]{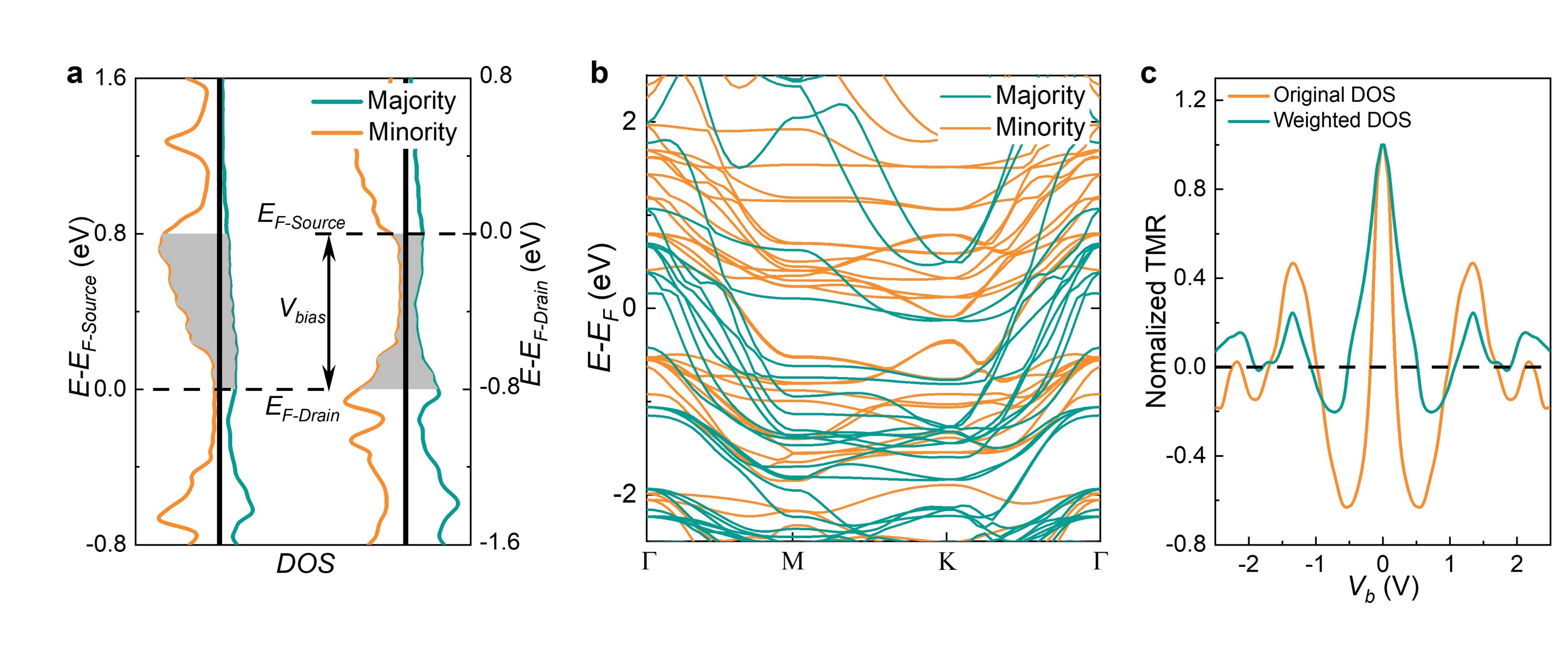}
\caption{\textbf{Theoretical model for inversion of TMR at different bias voltage.} \textbf{a} Tunneling process with high-energy electrons. \textbf{b} Calculated band structure of Fe$_3$GaTe$_2$. \textbf{c} Calculated TMR at different bias voltages. } 
\end{figure*}

The most plausible mechanism for TMR inversion is the involvement of high-energy electrons in the tunneling process, as illustrated in Fig. 4a. Since the DOS for majority and minority spins differs at various energy levels, the inclusion of higher-energy electrons enables tunable spin injection and detection \cite{TMRDOS1999PRL,TMRDOS1999}, leading to the observed TMR sign changes. To gain deeper insights into this mechanism, we calculated the band structure of FGT using density functional theory (DFT), and the results are shown in Fig. 4b and similar as previous calculations \cite{DFT2023APL}. From these calculations, we extracted the spin-resolved DOS at different energies. Assuming the tunneling process is elastic and neglecting the in-plane momentum conservation, the tunneling current can be expressed as \cite{FGTNC2023}, 

\begin{widetext}
\begin{equation}
\left\{\begin{array}{l}
I_{P} \propto \int_{\mu_{D}}^{\mu_{S}}\left[\rho_{D}^{\uparrow}\left(E-\mu_{D}\right) \rho_{S}^{\uparrow}\left(E-\mu_{S}\right)+\rho_{D}^{\downarrow}\left(E-\mu_{D}\right) \rho_{S}^{\downarrow}\left(E-\mu_{S}\right)\right] \mathrm{d} E \\
I_{A P} \propto \int_{\mu_{D}}^{\mu_{S}}\left[\rho_{D}^{\uparrow}\left(E-\mu_{D}\right) \rho_{S}^{\downarrow}\left(E-\mu_{S}\right)+\rho_{D}^{\downarrow}\left(E-\mu_{D}\right) \rho_{S}^{\uparrow}\left(E-\mu_{S}\right)\right] \mathrm{d} E
\end{array}\right.
\end{equation}
\end{widetext}
where $\rho_{S(D)}$ denotes the spin-resolved DOS of source (drain) electrode and superscript ↑(↓) represents majority (minority) spins. $\mu_{S(D)}$ is the chemical potential of source (drain) electrode and satisfies, $\mu_D-\mu_S=eV_b$ . We computed the bias voltage dependence of the TMR, which is defined as,
$TMR=I_P/I_{AP}-1$ .The calculation result is shown as the \comment{orange} line in Fig. 4\comment{c}. The theoretical results successfully reproduce the TMR inversions observed in our experiments.

While the theoretical model qualitatively agrees with the experimental observations, there remains a discrepancy in the exact bias voltage at which TMR inversion occurs. Experimentally, the TMR switches from positive to negative at \comment{0.625±0.025} V, whereas the theoretical prediction places this inversion at approximately 0.19 V. A similar discrepancy—where the experimental values exceed theoretical predictions—was also noticed in previous studies on MTJs using Fe$_3$GeTe$_2$ electrodes. The difference between experimental and theoretical results is particularly pronounced when GaSe is used as the tunneling barrier \cite{FGTNC2023} and less significant, though still present, in systems with WSe$_2$ or hBN barriers \cite{FGTNM2022}, even when more sophisticated models are applied.

The primary cause of this mismatch likely stems from the \comment{ignorance} of in-plane momentum conservation in MTJ with GaSe, even though the proposed theoretical model correctly captures the underlying physics. For MTJs with WSe$_2$ or hBN, the mismatch would be because of the simplified assumption that the crystals of the materials constituting the MTJ are perfectly aligned, to account for the in-plane momentum conservation during the tunneling process. This assumption holds well for MTJs fabricated using conventional thin films deposited by sputtering or molecular beam epitaxy, where crystal alignment is crucial for achieving high-quality MTJs. However, MTJs based on 2D materials do not require lattice matching and alignment. In fact, none of the existing experimental works on 2D MTJs report intentional crystal alignment yet. As a result, how to keep in-plane momentum conservation should be considered for calculations with misaligned electrodes. However, accurately calculating the TMR in MTJs with misaligned electrodes using ab-initio methods is challenging due to the large number of atoms involved in a unit cell.

To improve the theoretical results based on the Ref [15], we proposed a simplified approach by assigning different weights to the DOS at different k-points. In our MTJ, which uses hBN as the tunneling barrier, the barrier height remains largely the same for different momentum states (k-points), unless the applied bias voltage becomes excessively large. This allows us to disregard the misalignment effects of hBN. During tunneling process, electrons at the $\Gamma$ point automatically conserve in-plane momentum, resulting in the highest tunneling weight of DOS. However, for electrons away from the $\Gamma$ point, either elastic or inelastic scattering processes are required, leading to reduced tunneling weight of DOS. To account for this difference in weight of DOS, we introduced a weighting function $W(k)=\delta/\sqrt{\delta ^2+k^2}$  to describe the variation in tunneling efficiency at different k-points. The results of our revised calculations \comment{ (with $\delta =0.1\times2\pi/\text{\AA}$ )},  shown as the green line in Fig. 4\comment{c}, indicate that the voltage at which the TMR switches from positive to negative is now approximately 0.52 V, much closer to the experimental data. However, the voltage at which the TMR switches back to positive still differs significantly from the experimental value. This could be due to the fact that at higher bias voltages, the misalignment of the hBN barrier may also play a role and should be considered in future calculations.
                
\section*{Conclusions}

In conclusion, we have successfully demonstrated the bias voltage-controlled inversions of TMR in vdW heterostructures composed of FGT/hBN/FGT at room temperature. Notably, polarity reversal was observed consistently at \comment{0.625±0.025} V across various devices and temperatures. Our theoretical model, which incorporates spin-resolved DOS at high energy, shows qualitative agreement with the experimental findings. By adjusting the weight of the DOS at different k-points to account for electrodes misalignment, we significantly reduced the discrepancy between theory and experiment, bringing the inversion voltage closer to observed values. These findings enhance our understanding of the mechanisms behind voltage-controlled spin injection and detection, underscoring the promise of two-dimensional material-based MTJs for energy-efficient spintronic devices. Future work should focus on refining theoretical models to better account for misalignment effects and experimentally aligning different angles between ferromagnetic electrodes. Additionally, exploring the dependence of TMR inversion on voltage and its practical applications in next-generation memory and logic devices will be crucial.



\section*{Acknowledgement}
We thank Miss Manman Zhang at State Key Laboratory for Manufacturing Systems Engineering of Xi'an
Jiaotong University for her technical support. This work is financially supported by National Natural Science Foundation of China (Grants no. 12374121, 12304232 and 62274087), Shaanxi Fundamental Science Research Project for Mathematics and Physics (22JSY026,23JSQ011), Anhui Provincial Natural Science Foundation (Grants no. 2408085J025), the Anhui Provinical Major S\&T Project (Grants no.s202305a12020005) and the Fundamental Research Funds for the Central Universities.

\bibliographystyle{mynaturemag}
\bibliography{biblio.bib}
\end{document}


\title{Supplementary Materials for ``\textit{Bias Voltage Controlled Inversions of Tunneling Magnetoresistance in van der Waals Heterostructures 
Fe$_3$GaTe$_2$/hBN/Fe$_3$GaTe$_2$}"}


\author{Lihao Zhang}
\affiliation{\MOE}
\author{Miao He}
\affiliation{\HF}
\affiliation{\HFF}
\author{Xiaoyu Wang}
\author{Haodong Zhang}
\affiliation{\MOE}
\author{Keying Han}
\affiliation{\QD}
\affiliation{\Yanshan}
\author{Yonglai Liu}
\affiliation{\HF}
\affiliation{\HFF}
\author{Lei Zhang}
\affiliation{\MOE}
\author{Yingchun Cheng}
\affiliation{\QD}
\affiliation{\Yanshan}
\author{Jie Pan}
\affiliation{\MOE}
\author{Zhe Qu}
\affiliation{\HF}
\affiliation{\HFF}
\author{Zhe Wang}
\email{zhe.wang@xjtu.edu.cn}
\affiliation{\MOE}

\maketitle

\renewcommand\thesection{\arabic{section}}
\titleformat{\section}[block]
  {\normalfont\large\bfseries}
  {Supplementary Note \thesection:}{0.5em}{\centering}
\makeatother

\newcommand\4{$_4$}
\renewcommand\a{{\bf a}}
\renewcommand\b{{\bf b}}
\renewcommand\c{{\bf c}}
\renewcommand\d{{\bf d}}
\renewcommand{\theequation}{S\arabic{equation}}

\titleformat{\section}[block]
  {\normalfont\large\bfseries}
  {Supplementary Note \thesection:}{0.5em}{\centering}

\section{Theoretical model for TMR inversions under different bias voltage} 

The tunneling magnetoresistance (TMR) is related to the density of states (DOS) of ferromagnetic metal. For the density of states and band structure of Fe$_3$GaTe$_2$ crystal, the density functional theory (DFT)-based first-principles calculations were implemented in the Vienna Ab initio Simulation Package (VASP). During all calculations of study, we used the Perdew-Burke-Ernzerhof exchange-correlation functional within the generalized gradient approximation (GGA). Experimentally measured Fe$_3$GaTe$_2$ lattice constants of $a$ = 3.986 \AA, $b$ = 3.986 \AA, $c$ = 16.229 \AA, were assumed in the calculations. For relaxation and self-consistent calculations, a Monkhorst-Pack k-points mesh of $30\times30\times6$ was adopted, while we selected a k-points grid of $45\times45\times9$ for the calculation of density of states. During the calculations, the energy convergence criterion was set to 10$^{-6}$ eV. The DOS can be evaluated by,
\begin{equation}
    \rho(\varepsilon)=\frac{\Omega}{(2\pi)^3}\sum\int delta(\varepsilon-E(k)) d^3k.
\label{dos1}
\end{equation}
where $\Omega$ denotes the volume of unit cell, $delta$ represents the Delta function, the summation extends over all the bands.

\begin{figure*}
\centering
\includegraphics[width =0.5\linewidth]{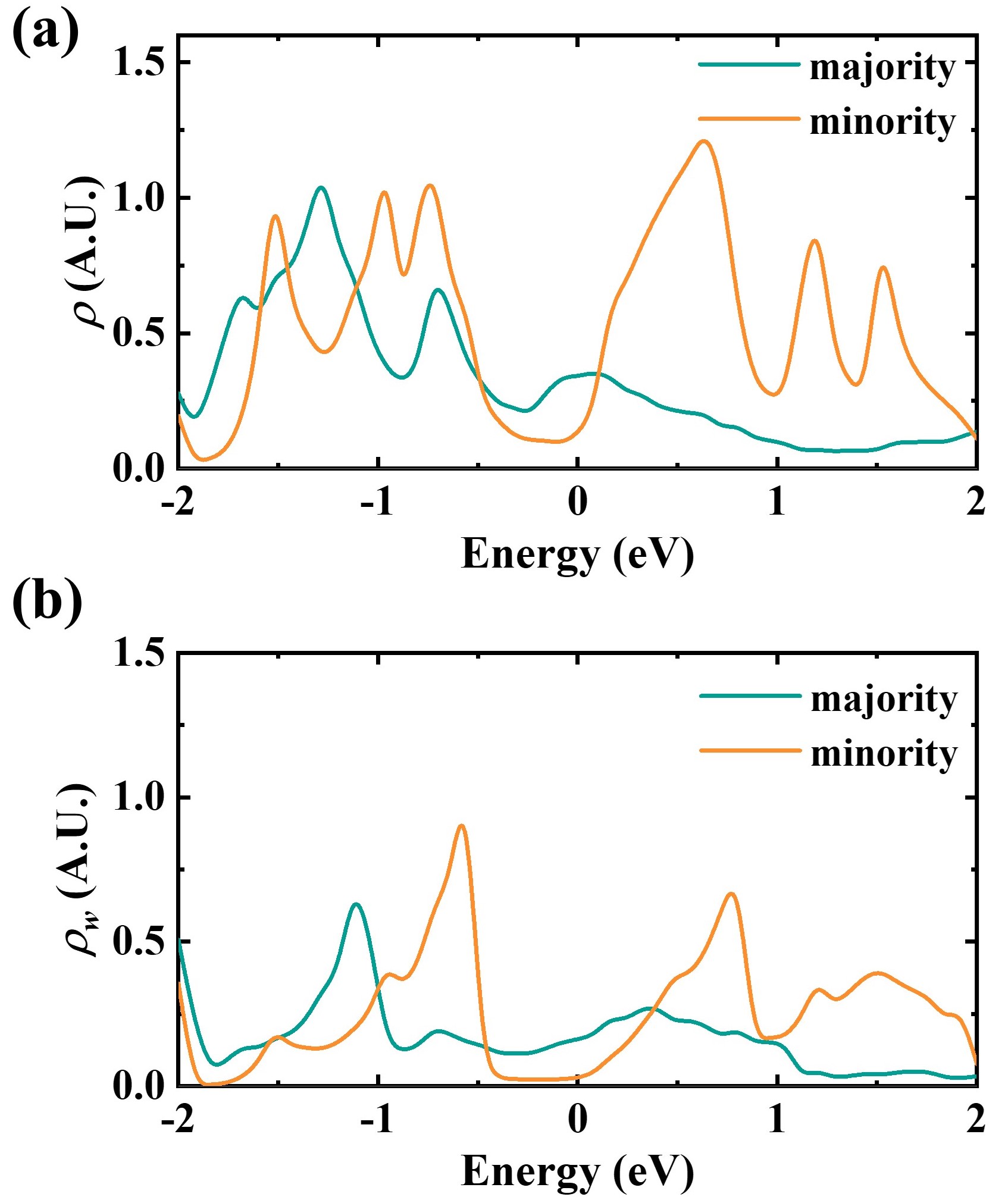}
\caption{(a) Original and (b) weighted DOS for Fe$_3$GaTe$_2$.} 
\end{figure*}

The tunneling current is proportional to the DOS of both Fe$_3$GaTe$_2$ electrodes. At different bias voltages, the energy window of participating electrons varies, leading to different spin polarizations. Considering energy conservation during the tunneling process, we have:
\begin{equation}
\left\{\begin{array}{l}
I_{P} \propto \int_{\mu_{D}}^{\mu_{S}}\left[\rho_{D}^{\uparrow}\left(E-\mu_{D}\right) \rho_{S}^{\uparrow}\left(E-\mu_{S}\right)+\rho_{D}^{\downarrow}\left(E-\mu_{D}\right) \rho_{S}^{\downarrow}\left(E-\mu_{S}\right)\right] \mathrm{d} E \\
I_{A P} \propto \int_{\mu_{D}}^{\mu_{S}}\left[\rho_{D}^{\uparrow}\left(E-\mu_{D}\right) \rho_{S}^{\downarrow}\left(E-\mu_{S}\right)+\rho_{D}^{\downarrow}\left(E-\mu_{D}\right) \rho_{S}^{\uparrow}\left(E-\mu_{S}\right)\right] \mathrm{d} E
\end{array},\right
\label{ip}
\end{equation}
where we set the chemical potential of the drain $\mu_D = 0$ and that of the source $\mu_S = -eV_b$, where $V_b$ is the bias voltage. $\rho_{S(D)}$ denotes the DOS of source(drain) electrode. The TMR is then defined as,
\begin{equation}
    TMR=I_P/I_{AP}-1
\label{TMR}
\end{equation}
By substituting Eq. \ref{ip} into Eq. \ref{TMR}, we derive the TMR as a function of the bias voltage, The calculated results, presented in Fig. 4(c) of the main text, reveal notable TMR sign reversals. These reversals highlight the dependence of spin polarization on the energy window defined by $V_b$, which is directly influenced by the density of states (DOS) of the source and drain electrodes. 

The calculations above consider only energy conservation, neglecting in-plane momentum conservation during the tunneling process. When there is a lattice orientation mismatch between the bottom and top Fe$_3$GaTe$_2$ layers, the momentum of an electron at $\boldsymbol{k}$ point in one layer must be scattered into $\boldsymbol{k'}$ in the other. The momentum difference $\boldsymbol{\Delta k}=\boldsymbol{k}-\boldsymbol{k'}$ arises from scatterings caused by an external potential $V$. The scattering probability is proportional to $<\boldsymbol{k}|V|\boldsymbol{k'}>$. Assuming $V$ varies slowly, the scattering probability $p$ can be approximated as $p \propto \int \exp(i \Delta k r) dr \propto 1/\Delta k \propto 1/k$. This relation suggests that the scattering probability is maximized at the $\Gamma$ point. To incorporate this effect, we compute the weighting function that quantifies the relative contributions of different $k$ values in the tunneling process.
\begin{equation}
    W(k)=\delta/\sqrt{\delta^2+k^2}.
\label{weight}
\end{equation}
This weighting function serves to adjust the tunneling probabilities to reflect momentum conservation constraints more accurately. With this weighting function, the density of states (DOS) can be re-evaluated as:
\begin{equation}
    \rho_w(\varepsilon)=\frac{\Omega}{(2\pi)^3}\sum\int W(k) delta(\varepsilon-E(k)) d^3k.
\label{dos2}
\end{equation}
By setting $\delta =0.1\times2\pi/\text{\AA}$, we calculate the weighted density of states (DOS), $\rho_w$,  incorporating the momentum-dependent scattering probabilities as shown in Supplementary Fig. 1(b). Comparing Supplementary Figs. 1(a) and 1(b), we observe a significant change: the minority DOS surpasses the majority DOS at higher energy levels for the weighted DOS. This shift highlights the impact of momentum conservation during tunneling, emphasizing that states with lower momentum mismatch contribute more prominently to the transport properties.

By substituting Eq. \ref{dos2} into Eqs. \ref{ip} and \ref{TMR}, we calculate the tunneling magnetoresistance (TMR) using the weighted DOS. The results, represented by the green curve in Fig. 4(c) of the main text, reveal how momentum-dependent scattering probabilities influence the TMR behavior, offering a refined understanding of its voltage-dependent characteristics.